\documentclass[aps,pre,twocolumn,showpacs,superscriptaddress,groupedaddress]{revtex4-2}
\usepackage{amsmath}
\usepackage{amsfonts}
\usepackage{amssymb}
\usepackage{graphicx}
\usepackage{graphics}
\usepackage{dcolumn}
\usepackage{sidecap}
\usepackage{parskip}
\usepackage{color}
\usepackage{hyperref}
\usepackage{hhline}
\usepackage{mathtools}
\usepackage{multirow}
\usepackage{verbatim}
\usepackage{rotating}
\usepackage{setspace}
\usepackage{epsfig}
\usepackage{epstopdf}
\usepackage{subfigure}
\usepackage{booktabs}
\usepackage[normalem]{ulem}
\usepackage{bm}
\usepackage{verbatim}
\usepackage{comment}
\usepackage{cleveref}
\usepackage[normalem]{ulem}

\makeatletter
\def\@eqnnum{{\normalsize \normalcolor (\theequation)}}
 \makeatother

\graphicspath{{./}{ER/main/}}
\begin{document}
\title{Oscillation Quenching in Stuart-Landau Oscillators via Dissimilar Repulsive Coupling}
\author{Subhasanket Dutta$^{1}$, Omar Alamoudi$^{2,4}$, Yash Shashank Vakilna$^{2}$, Sandipan Pati$^{2}$} 
\email{patilabuab@gmail.com} 
\author{Sarika Jalan$^{1,3}$}\email{Corresponding Author: sarika@iiti.ac.in}
\affiliation{1. Complex Systems Lab, Department of Physics, Indian Institute of Technology Indore, Khandwa Road, Simrol, Indore-453552, India}
\affiliation{2. Texas Institute for Restorative Neurotechnologies, The University of Texas Health Science Center at Houston, Houston, TX-77225, USA}
\affiliation{3. Department of Biosciences and Biomedical Engineering, Indian Institute of Technology Indore, Khandwa Road, Simrol, Indore-453552, India}

\affiliation{4. Biomedical Engineering program, King Abdulaziz University, Jeddah Saudi Arabia}

\date{\today}

\begin{abstract}
	Quenching of oscillations, namely amplitude and oscillations death, is an emerging phenomenon exhibited by many real-world complex systems.
	Here, we introduce a scheme that combines dissimilar couplings and repulsive feedback links for the interactions of Stuart Landau oscillators and analytically derive the conditions required for the amplitude death.
	Importantly, this analysis is independent of the network size, presents a generalized approach to calculate the stability conditions for various different coupling schemes, and befits for non-identical oscillators as well.
	Last, we discuss the similarities of the quenching of oscillations phenomenon with the postictal generalized EEG suppression in convulsive seizures.

\end{abstract}


\maketitle

\section{Introduction}
Coupled Stuart Landau (SL) oscillators can display a wide range of emerging phenomena, such as synchronization, pattern formation, quenching of oscillations, etc. Especially quenching of oscillations  arising due to coupling between the pairs of oscillators has drawn considerable attention from the nonlinear dynamics
community due to  the widespread occurrence   of this phenomenon 
in many natural systems. For example, in laser systems, various types of couplings among the laser components can lead to the quenching of oscillation \cite{laser}. In neurological systems, oscillation death has been proposed to be the root cause of various neurodegenerative diseases and has been modeled using coupled nonlinear oscillators \cite{prasad_review}. A few other systems manifesting quenching of oscillations are atmosphere \cite{climate}, electronic circuits \cite{banerjee_exp}, etc. 

There have been persistent efforts to realize the quenching of oscillations in various nonlinear model systems, among which coupled Stuart-Landau oscillators turn out to be apt for understanding the origin and implications of such behaviors. Quenching of oscillations in coupled Stuart Landau oscillators are primarily achieved in three manners, by introducing a parameter mismatch \cite{koseka}, communication delays \cite{koseka_review}, and conjugate coupling \cite{chaos_2021}.
Few experimental setups incorporating these designs have successfully achieved oscillations quenching or the death state \cite{banerjee}. A death state of an oscillator can be classified into two major categories, amplitude death (AD) and oscillation death (OD).  Amplitude death corresponds to all the oscillators settling down to the same fixed point located at the origin. In oscillation death, oscillators settle at different fixed points or the same fixed point away from the origin \cite{koseka_review}. 
The dynamical equation for an uncoupled SL oscillator can be written by,
\begin{equation}
\label{uncoupled_sl}
	\dot z(t)=(a^2-|z(t)|^2)z+i\omega z
\end{equation}
Here $z$ is a complex variable depicting the dynamical state of an oscillator with $\omega$ being its intrinsic frequency. The oscillator has one unstable fixed point acting as a center for a stable circular limit cycle of radius $a$.
 
A set of identical SL oscillators ($\omega_i=\omega_j$) is unable to show quenching of oscillation with a simple diffusive coupling through the $z$ variable. However, different coupling schemes play different governing roles in determining steady-state behaviors. For instance, the oscillation death state is achieved for the coupling term being present only in the real or imaginary part of $z$ \cite{koseka}. In fact, there could be a transition from the oscillatory state (OS) to AD when $z$ is coupled diffusively with its conjugate $z^*$ \cite{chaos_2021}. Moreover, in the identical oscillators, repulsive feedback coupling \cite{hens_pal_dana,nandan}  and diffusive coupling in dissimilar (also referred to as conjugate) variables \cite{prasad_diss_1,prasad_review} can steer AD and OD. In non-identical oscillators,  frequency mismatch and coupling strength are enough to bring the oscillator death, even for simple diffusive coupling via the $z$ variable.
Non-identical SL oscillators coupled through diffusive couplings on
small-world networks have shown to support OD \cite{hou2003,rubchinsky2000}, and on scale-free networks have shown to yield a complete AD state \cite{Liu_2009}.
Lately, the first-order abrupt transition to AD, popularly referred to as explosive death, has become a topic of great interest due to the theoretical curiosity fueled by observations of the phenomenon's existence in many real-world systems. 
The first-order transition to AD can be successfully induced in coupled oscillators via an environmental coupling scheme in a single layer \cite{verma1,verma2} as well as in multiplex networks  \cite{verma3,multiplex}.

Further, there have been persistent efforts to model large-scale brain networks using coupled oscillators on networks \cite{kura_brain,unlee,postic_1,postic_2}. 
For example, few previous studies have considered simple linear phase oscillators to understand various emerging dynamical features of brain networks \cite{kura_brain}. 
A simple brain network model consists of firing neurons acting as nodes connected through synapses defining interactions between the pairs of neurons.  Dynamical behaviors of an individual neuron/node were first studied by the  Hodgkin-Huxley model, explaining the initiation and propagation of action potential in neurons \cite{hodgkin}.   Subsequently, other models like Fitzhugh-Nagumo (explaining spiking in neurons) and Hindmarsh-Rose (explaining spiking-bursting in neurons) \cite{fitzhugh,hindmarsh} were discovered. However, in large-scale brain networks, the collective behavior of the nodes was considered to be quite low-dimensional, and information about the dynamical evolution of each neuron was shown to be relatively irrelevant \cite{brain_review,pathak_roy_ban}. A well-known model representing large-scale brain dynamics is the neuronal mass model, which presents an ensemble approach in which the dynamics of a patch of the cortex (a local population of neurons) are represented by a set of differential equations  reduced in dimension \cite{brain_review}. Further advances in this field led to the discovery of more realistic models, such as the whole brain model and the brain networks model, where brain areas were treated as nodes in a coupled dynamical system.  These patches of the cortex are considered as units or nodes coupled with each other according to their anatomical connectivity  (edge) patterns. 
Simplified models like the Kuramoto oscillator \cite{kura_brain_review} and SL oscillators \cite{sl_eg_1,sl_eg_2,unlee} have been used in a similar fashion to portray large-scale brain dynamics. 
Different models are constructed such that they can explain the phenomenon of interest \cite{pathak_roy_ban,kura_brain_review}. 

Synchronized activities among the different brain regions have been associated with the onset of the seizure from the pre-seizure region \cite{kura_brain}. It is common knowledge that soon after the generalized tonic-clonic seizures, the brain state exhibits a transition to an isoelectric EEG state, with the existence of a profound scalp EEG voltage attenuation ($<10$ microvolts), referred to as the
postictal generalized EEG suppression (PGES) \cite{postic_1,postic_2}. While we do not claim that the model presented here provides a mechanism behind the occurrence of PGES in the human brain, the phenomenon depicted by this model bears a close resemblance to PGES. Furthermore, to make the modelling of PGES more realistic, for the coupling matrix of Eq. 1, we have considered the functional correlation matrices generated for the EEG time series data from brain during seizure.

First, we develop a  theoretical framework to analyze amplitude death in coupled SL oscillators on complex networks. Earlier theoretical works on SL oscillators pertain to the linear stability analysis for direct mean-field diffusive coupling  \cite{strogatz_ad} and diffusive conjugate coupling \cite{chaos_2021} on globally coupled and star networks \cite{frasca}.  We consider  SL oscillators with dissimilar repulsive feedback couplings and develop
an analytical approach that is independent of the size of the underlying coupling network. The method is a generalized one as it facilitates calculations of necessary and sufficient conditions to attain amplitude death for other coupling forms. The key lies in the fact that the analysis uses a generalized form of coupling matrix, providing it an edge over previously existing frameworks.
Then, we numerically study the dynamical behaviors of this setup on various network architectures, namely,  globally coupled networks, regular lattice networks, Erd\"os-Reny\'i (ER) random networks.
Finally, we numerically analyze the results of Eq.~1 with the functional coupling matrices generated from real time series EEG data of patients, and discuss the similarity between the phenomenon depicted by the model with PGES.

\section{Model}\label{sec:Models}
Dynamical evolution of an uncoupled SL oscillator is governed by Eq.~\ref{uncoupled_sl}. Upon substituting $z=x+iy$, the resulting equation is,
\begin{equation}\label{eqn:uncoupled_short} 
\dot x_k= P^x_k,\: \: \dot y_k=P^y_k 
\end{equation}
where,
\begin{equation}
P^x_k=(1-x_k^2-y_k^2)x_k-w_k x_k,\: \: P^y_k=(1-x_k^2-y_k^2)y_k+w_k y_k
\nonumber
\end{equation}
An introduction of the dissimilar repulsive feedback coupling between a pair of connected nodes through both the $x$ and $y$ coordinates results in the following equation, 
\begin{equation}\label{eqn:SL_diss_xy} 
\dot x_k= P^x_k
	-\frac{\varepsilon_x}{N}  \sum_{j=1}^{N}A_{jk}(y_j+x_k),\\ \dot y_k=P^y_k -\frac{\varepsilon_y}{N}  \sum_{j=1}^{N}A_{jk}(x_j+y_k)
\end{equation}
Here $A_{jk}$ is the adjacency matrix representing the underlying network structure. For an unweighted network, its elements take the value 1 when $j^{th}$ and $k^{th}$ nodes are connected, and 0 otherwise. Whereas, for a weighted network, the elements of the adjacency matrix are represented by the interaction weights.
For identical oscillators, it has been found that an introduction of the similar diffusive coupling does not yield oscillator death(see supplementary material \cite{SM}), whereas dissimilar repulsive feedback links bring oscillator death. 
Further, we consider different cases with the nodes having different coupling schemes and couplings being in only one dimension. We consider the weighted interaction matrix generated from the post-seizure data. First, we model the eight intracranial channels of the brain as non-identical Stuart-Landau oscillators on a directed weighted network. The dynamical equation of such nodes can be given by Eq.~\ref{eqn:SL_diss_xy} with  $\omega$  chosen from a Lorentzian frequency distribution with mean and standard deviation corresponding to the frequency band. Further, we study the dynamics of each node and the network as a whole.
 
\begin{figure}[t!]
    \centering
    	\includegraphics[height=5cm,width=\linewidth]{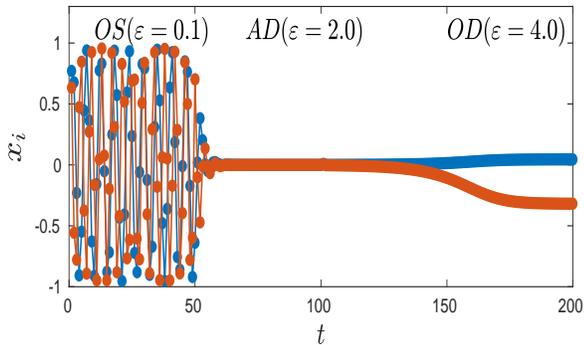}\\ 
    	\caption{$x_i$ vs $t$ plot depicting three different dynamical states for two randomly selected nodes form a network of $N=1000$ (Eq.~\ref{eqn:SL_diss_xy}). Lorentzian frequency distribution parameters are $\omega_0=2.0$ and $\Delta=0.30$.}
    	\label{fig:timeseries}
\end{figure}
Coupling of a SL oscillator with another SL oscillator  may either give birth to additional fixed points than those existed for the single uncoupled oscillator, or  may change stability properties of the existing fixed point(s).
In the following we define an order parameter which quantifies the variance of fluctuation of the dynamical variables over a time span, which tends to $0$ for both OD and AD cases:
\begin{equation} \label{eqn:order1}
    r= \frac{1}{N}\sum_{i=1}^{N} \langle x_i \rangle _{max,t}- \langle x_i \rangle _{min,t}\
    E=\frac{\sum_{i=1}^{N}|z_i|^2}{N}
    \nonumber
\end{equation}
For the numerical purpose, if $0 < E < 0.001$ and $0 < r < 0.001$, we infer that the system has reached the state of AD.

  Further, the different network architectures to define  pair-wise interaction matrices are constructed as follows. In a regular 1-d ring lattice, each node is connected to its $k$ neighbors. To produce an ER Random network with an average degree $n/N$, we randomly connect $n$ edges out of the possible $N(N-1)/2$ edges \cite{barabasi_review}. The scale-free network here is created via the Barabasi-Albert model in the following manner; starting with an initial $m_o$ number of nodes, at each time step, one node with $m$ connections($m<m_o$) is added. The connection probability of this new node getting attached to an existing node is proportional to the degree of that node.

\begin{table}
	\centering
	\begin{tabular}{lll}
		\toprule
		\midrule 
		Name     & A      & E \\
		\midrule
		AD      &=0      & =0     	 \\
		OD     & =0    & $\neq 0$   	 \\
		OS     &$\neq 0$  &$\neq 0$  \\
		\bottomrule
	\end{tabular}
	\caption{Amplitude (A) and oscillation (E) death measures for different states.}
	\label{tab:table_1}
\end{table}

\section{Analytical Calculation}
Let us now provide the mathematical formalism to analyze the stability of coupled SL oscillators on various setups. 
First, we present generalized characteristic equation the coupled SL oscillators on globally coupled networks. Thereafter,  we consider globally coupled nodes having dissimilar repulsive feedback couplings and try to solve using the generalized characteristics equation. We further mix various kinds of couplings and study the dynamical evolution of the coupled oscillators for the mixed setup.\\
At first, let us consider a setup of globally coupled SL oscillators with a generalized form of the coupling matrix $F$, which can be changed later according to the different setups. The generalized coupled differential equations for globally coupled networks of the Stuart Landau oscillators can be written as,
\begin{equation}
\begin{bmatrix}
    \dot x_k\\
    \dot y_k
\end{bmatrix}
=
\begin{bmatrix}
    (1-x_k^2-y_k^2)x_k-i\omega_k y_k \\
    (1-x_k^2-y_k^2)x_k-i\omega_k y_k 
\end{bmatrix}
+F(x_j,x_k,y_j,y_k)
\nonumber
\end{equation}
 The corresponding  Jacobian matrix takes the form:
\begin{align}\label{jaco_general}
    |I\lambda- M| = \left(
\begin{array}{ccccc}
   \ M_1+F_1 & . & . & F_1 \\
   \ F_2 & M_2+F_2 & . & . \\
   \ F_i & . & M_i+F_i & . \\
   \ F_N & . & . & M_N+F_N \\
\end{array}
\right)
\nonumber
\end{align}
\\
where,
$M_i = 
\begin{pmatrix}
    \lambda-a_1 & -\omega_i\\
    \omega_i & \lambda-a_2
\end{pmatrix}
$.
However, $a_1$, $a_2$ and $F_i$ vary in accord with the coupling scheme of  each node.
To deduce  the stability conditions for the AD state,  the  challenge  lies in solving the characteristic equation for the corresponding Jacobian matrix ($M$) given by $det(I_{2N}\lambda-M)=0$. Let us first solve $det(I_{2N}\lambda-M)=0$, for which we use the following lemma.\\
\textit{Matrix-Determinant Lemma: If $X$  is  $n \times  n$,  and $U$ and $V$ are  $n \times m$ matrices, 
\begin{equation}\label{eqn:lemma}
  |X+UV^T|=|X| \times |I_m+V^T X^{-1} U|    
\end{equation}
where $I_m$ is an identity matrix of the dimension $m \times m$ and  $O_{n\times n}$ is a null matrix of dimension $n \times n$.
}\\
Proof:
\begin{equation}
|X+UV^T|=
\begin{vmatrix}
    X+UV^T & U\\
    O_{n\times n} & I_m
\end{vmatrix}
\begin{vmatrix}
    I_n & O_{n\times m}\\
    -V^T & I_m
\end{vmatrix}
=
\begin{vmatrix}
    X & U\\
    -V^T & I_m
\end{vmatrix}
\nonumber
\end{equation}
\begin{equation}
\begin{vmatrix}
    X & U\\
    -V^T & I_m
\end{vmatrix}
=|X|\times|I_2+V^TX^{-1}U|
\nonumber
\end{equation}
using identity for the determinant of block matrices,
\begin{equation}\label{eqn:block}
\begin{vmatrix}
    A & B\\
    C & D
\end{vmatrix}
=|D|\times|D-CA^{-1}B|    
\end{equation}\\
Next, we apply the following lemma to find out the characteristic equation and the corresponding eigenvalues.
$(\lambda I_{2N}-M)$ can be written as;
\begin{equation}
   (\lambda I_{2N}-M)=M_d + UV^T
   \nonumber
\end{equation}
where\\
$
M_d=
\begin{bmatrix}
   M_1 & 0 & 0  & . & . & . \\ 
   0 & M_2 & 0 & 0 &  . &  .\\ 
   . & . & . & . & . & . &\\
   . & . & . & . & . & . & \\
   0 & 0 & . &  .  &  .  & M_N\\ 
\end{bmatrix}, \, U=
\begin{bmatrix}
    F_1 \\ F_2 \\ F_3 \\ .\\ .\\
\end{bmatrix}, \,
V=\begin{bmatrix}
    I_2 \\ I_2 \\ I_2 \\ .\\ .\\
\end{bmatrix}
$ \\       

By using the matrix determinant lemma (Eq.~\ref{eqn:lemma}) and Eq.~\ref{eqn:block} we obtain, 
\begin{equation}
\begin{split}
    |&\lambda I_{N/2}-M|=|M_d|  \times |I_2+V^TM_d^{-1}U|\\
    &=|M_d|  \times |I_2+V^T(\frac{adj(M_d)}{|M_d|})U|\\
    &=\Pi_{i=1}^N|M_i|\times |I_2+\sum_{i=1}^{N}\frac{adj(M_i)F_i}{|M_i|} |\\
\end{split}
\nonumber
\end{equation}
The generalized characteristic equation are given by,
\begin{equation}\label{master}
    \Pi_{i=1}^N|M_i|=0\;\;\text{and}\;\;|I_2+\sum_{i=1}^{N}\frac{adj(M_i)F_i}{|M_i|} |=0
\end{equation}
where $adj(M_i)$ is the adjoint matrix of $M_i$. What follows that the characteristic equation reduces to a determinant of a $2 \times 2$ matrix which is independent of the size of the network. Moreover, such a reduction allows this scheme to work even when nodes are connected through a different form  of  the coupling provided each node has the same coupling scheme separately.
Let us consider various different cases as test-beds for the analysis. Note that, the simplest case is the one where identical oscillators are diffusively coupled. This coupling does not exhibit AD \cite{SM}.
\begin{figure}[t!]
    \centering
    \begingroup
    \setlength{\tabcolsep}{-6pt} 
    \renewcommand{\arraystretch}{-1.0}
    	\begin{tabular}{cccc}
    	\includegraphics[height=3.5cm,width=0.5\linewidth]{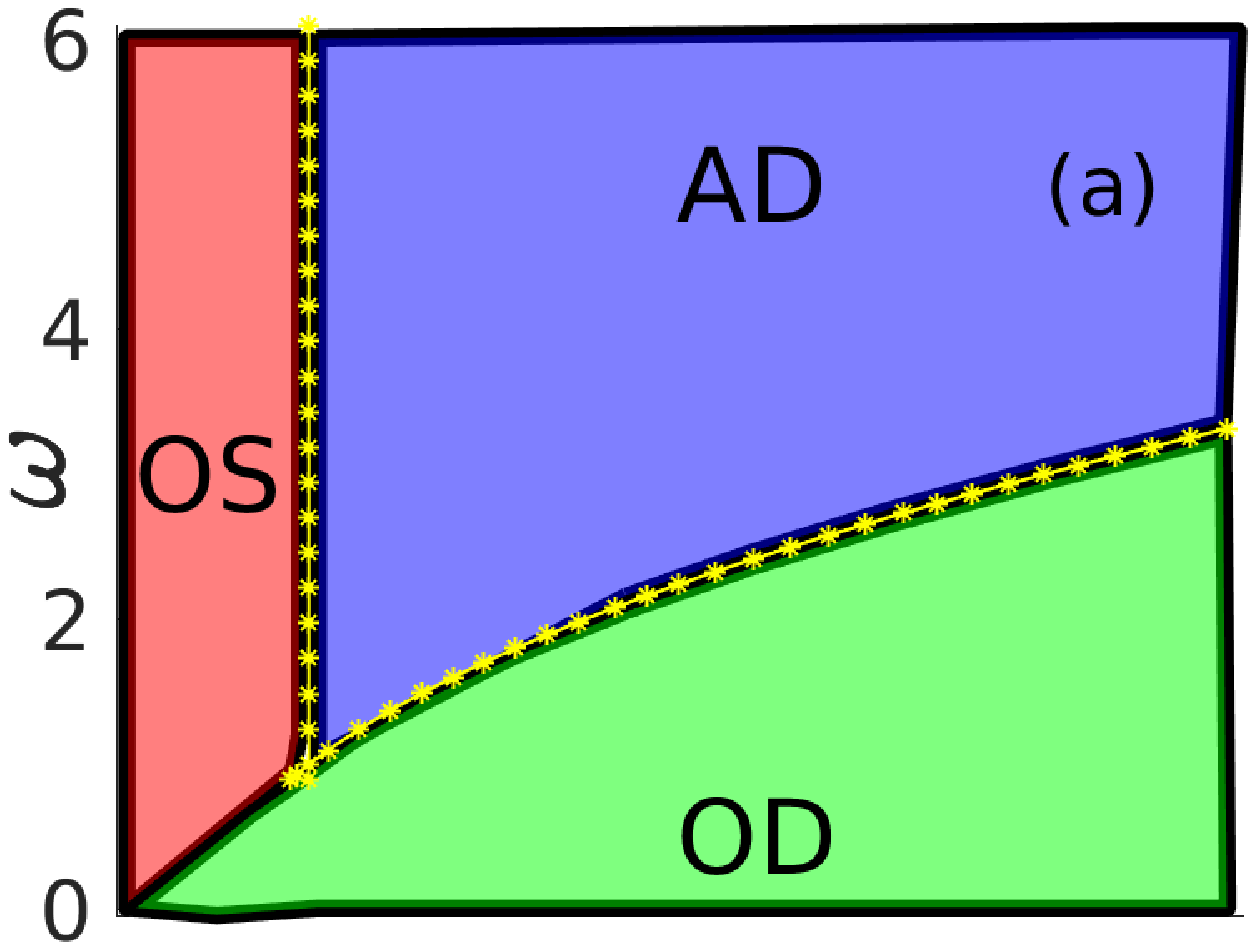}& 
    	\includegraphics[height=3.5cm,width=0.5\linewidth]{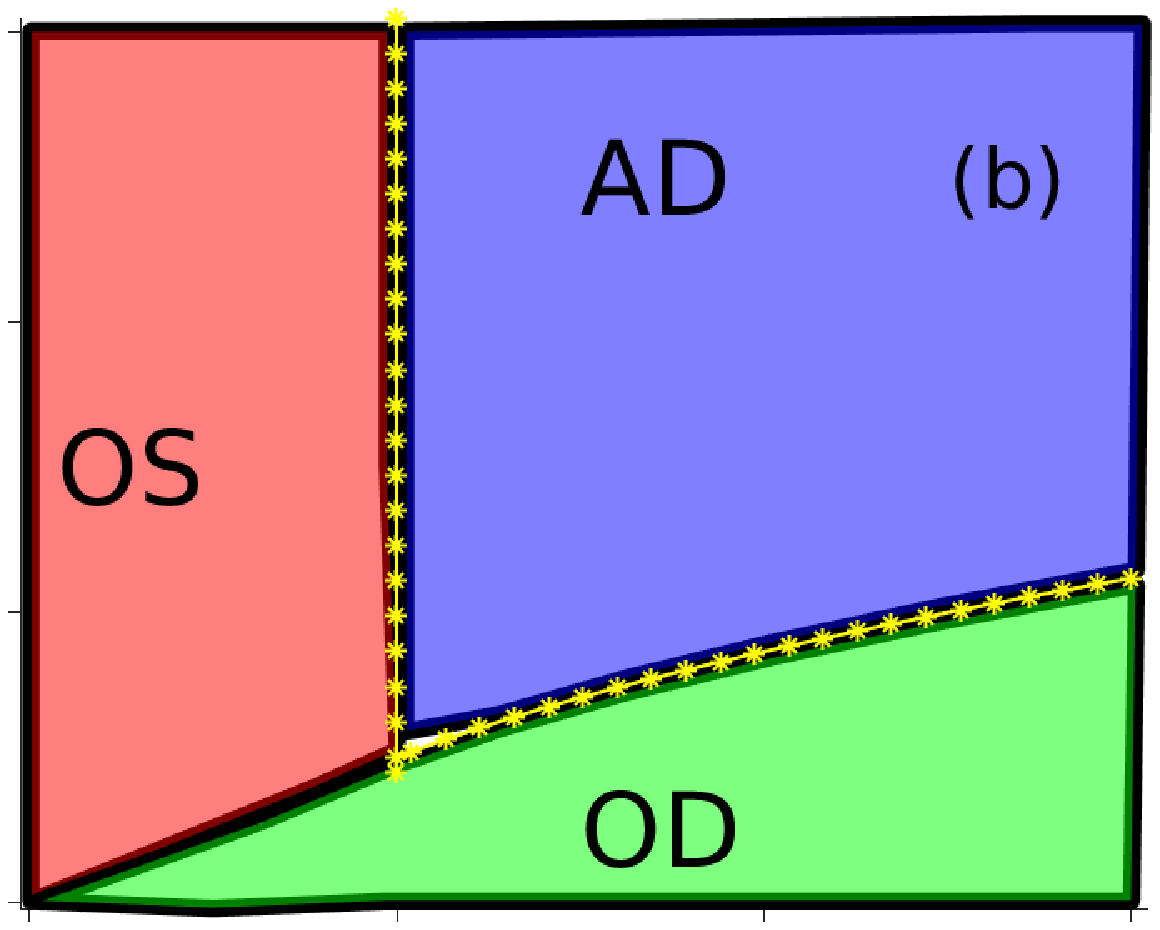}\\
    	\includegraphics[height=3.7cm,width=0.5\linewidth]{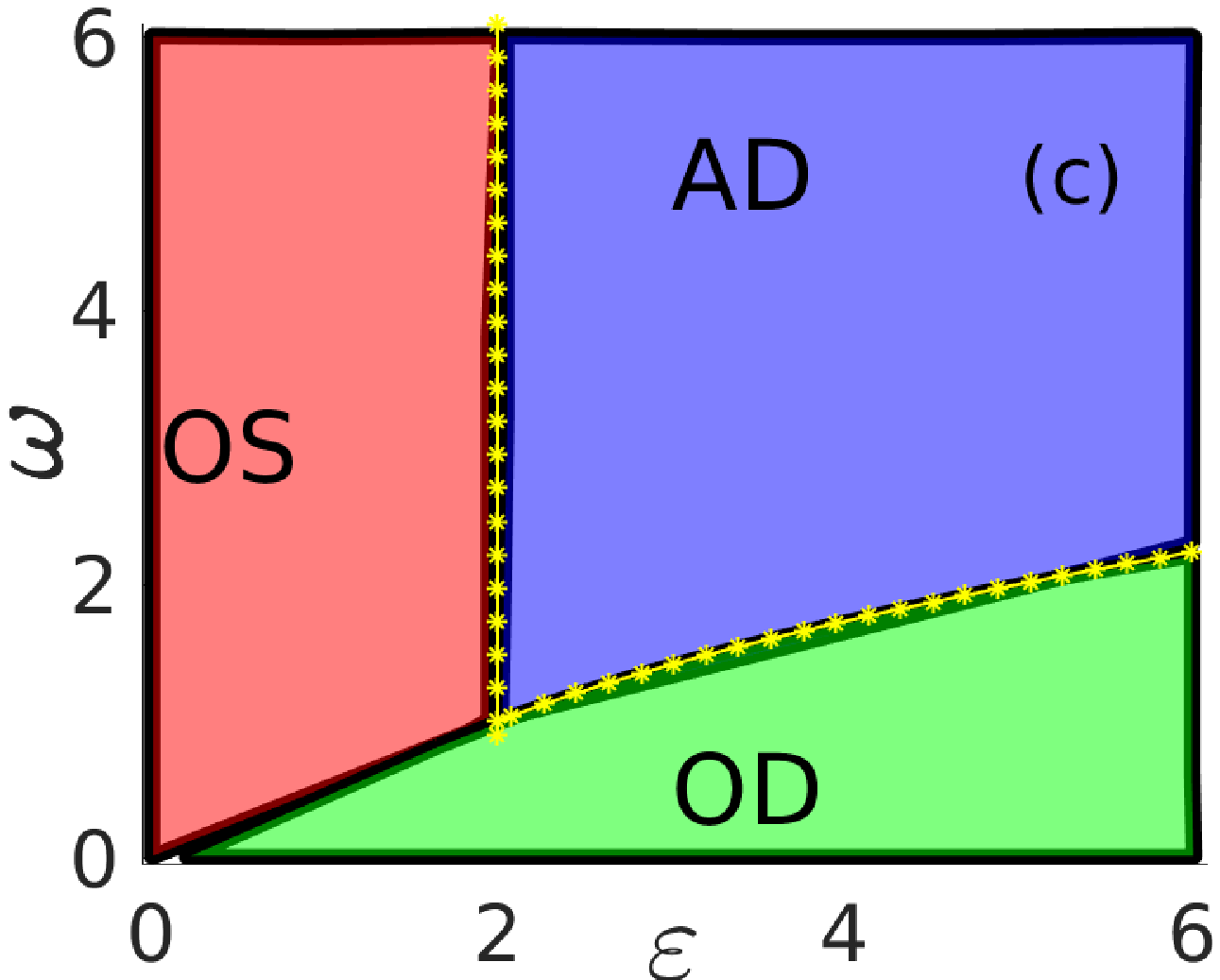}&  
    	\includegraphics[height=3.7cm,width=0.5\linewidth]{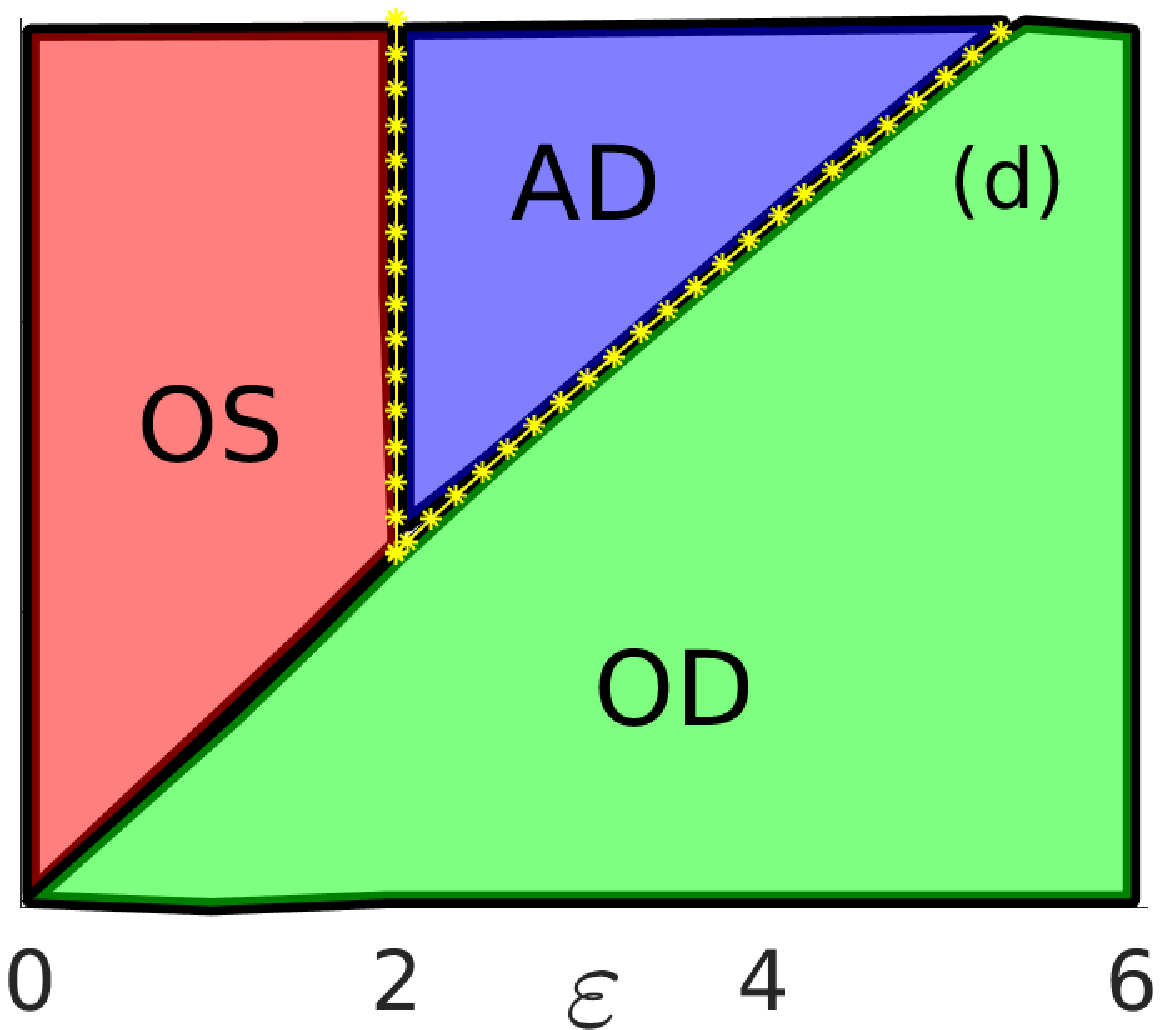}\\
    	\end{tabular}{}
    \endgroup
    	\caption{Parameter space plot $\omega$ vs $\varepsilon$ for  globally coupled network of size $N=1000$ consisting of identical oscillators ($\omega_i = \omega_j = \omega \forall i,j$). (a) all nodes having dissimilar repulsive feedback coupling, (b) with half of the nodes having dissimilar repulsive feedback and another half with direct diffusive couplings,
    	(c) dissimilar repulsive feedback coupling via $x$ variable only, (d) dissimilar repulsive feedback coupling via $y$ variable only. The solid black lines represent the transition boundaries calculated numerically, whereas the yellow line represents the analytical region corresponding to the stable origin (AD).}
    \label{fig:para_diagram}
\end{figure}

\begin{figure}[t!]
    \centering
    \centering
    \begingroup
    \setlength{\tabcolsep}{-2pt} 
    \renewcommand{\arraystretch}{0.0}
    \begin{tabular}{cc}
        \includegraphics[height=4.85cm,width=0.55\linewidth]{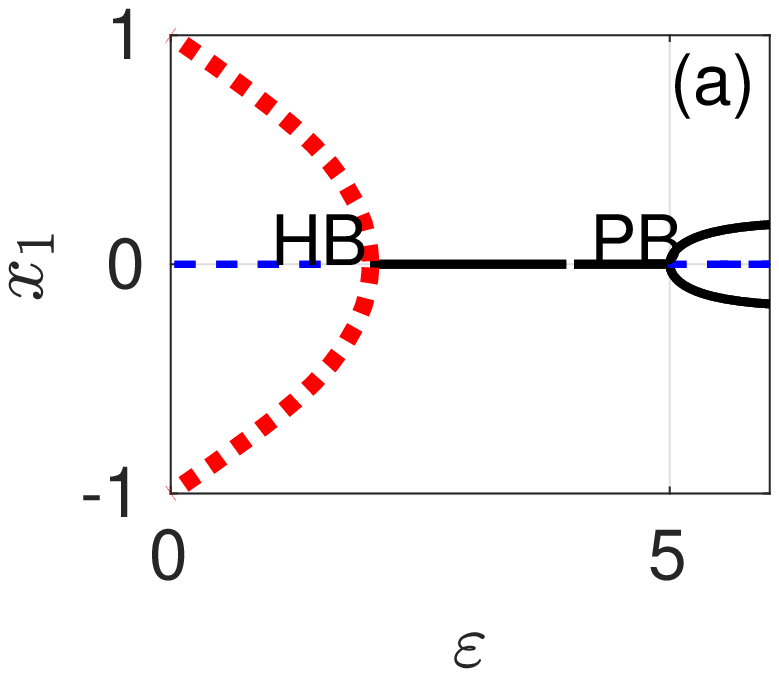}&
        \includegraphics[height=5.0cm,width=0.55\linewidth]{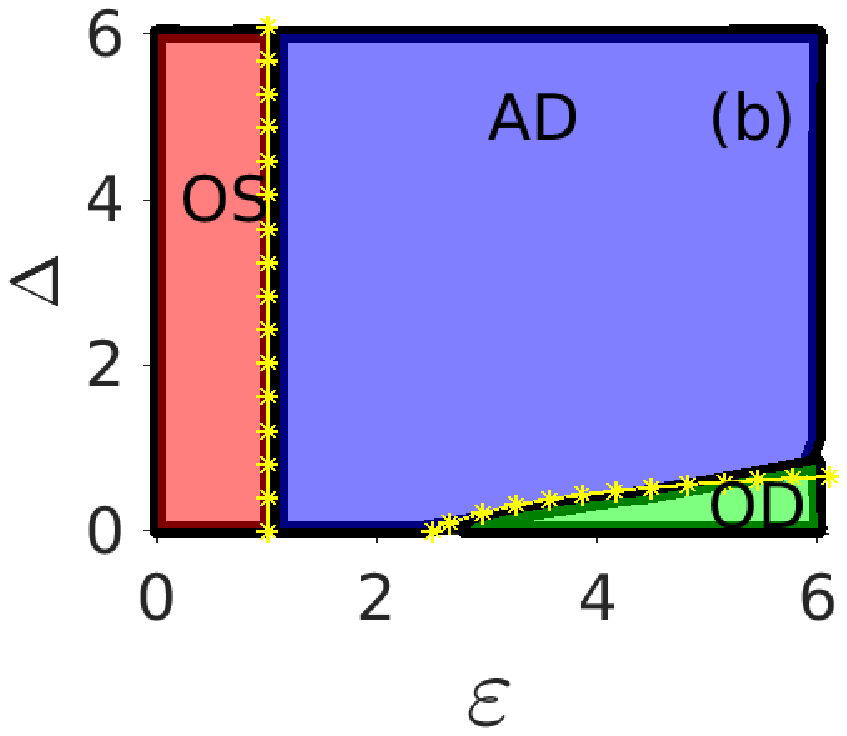}
        \end{tabular}
    \endgroup
    \caption{(a) Bifurcation plot for dissimilar repulsive coupling via $x$ variables for identical oscillators ( $\omega=2.0$), red dashed line corresponds to stable limit cycle, solid black line represents stable fixed point, and dashed black line represents an unstable limit cycle. (b) Parameter space $\Delta$ vs $\varepsilon$ plot for  non-identical globally coupled networks of size $N=1000$ and Lorentzian frequency distribution with $\omega_0=2.0$. The solid black and yellow dashed lines represent the transition boundaries calculated numerically and analytically, respectively.}
    \label{fig: bif and non}
\end{figure}

\paragraph{\textbf {Introduction of dissimilar repulsive feedback coupling:}}
First, we consider a system where all nodes are coupled via dissimilar repulsive feedback. 
The matrices $M_i$ and $F_i$ which remain the same for all values of $i$, can be given by,
$
M_i=
\begin{pmatrix}
    \lambda-1+\varepsilon & \omega\\
    -\omega & \lambda-1+\varepsilon
\end{pmatrix}
\text{and }    
F_i=
\begin{pmatrix}
    0 & \varepsilon/N \\
    \varepsilon/N &0
\end{pmatrix}
\text{and }    
$
This setup yields the following eigenvalues:
\begin{equation}\label{eqn:eig1}
    \lambda=1-\varepsilon \pm \sqrt{\varepsilon^2-\omega^2}, \:\: \lambda=1-\varepsilon \pm i\omega_o
\end{equation}
The conditions for the origin to be a stable state is satisfied for $Re[\lambda_i]<0$ for all $i$. Applying this condition, from Eq.~\ref{eqn:eig1} we get that for (i) $\varepsilon < \omega$, $\varepsilon > 1$,  and  (ii) $\varepsilon  > \omega$, $1-\varepsilon \pm \sqrt{\varepsilon^2-\omega} <  0$  yielding  $\varepsilon < (1+\omega^2)/2$.\\
Now we consider a system of identical oscillators with two kinds of coupling. Some nodes have coupled via repulsive dissimilar coupling, whereas others have coupled via simple diffusive coupling.
The dynamical equation for the simple diffusively coupled nodes can be given by Eq.~S1 of the Supplementary material \cite{SM}, whereas the nodes with the repulsive link will be governed  Eq.~\ref{eqn:SL_diss_xy}.
Proceeding similarly to the last section, the coupling matrices for the non-repulsive and the repulsive schemes are given by $F_1$ and $F_2$, respectively. However, the matrices $M_1$ and $M_2$, remain the same for both types of nodes. The $M_1$,$F_1$ and $F_2$ matrix are given by,
$
M_1=
\begin{pmatrix}
    \lambda-1+\varepsilon & \omega\\
    -\omega & \lambda-1+\varepsilon
\end{pmatrix}
\text{and }    
F_1=
\begin{pmatrix}
    -\varepsilon/N & 0\\
    0 & -\varepsilon/N
\end{pmatrix}
\text{and }    
F_2=
\begin{pmatrix}
    0 & \varepsilon/N \\
    \varepsilon/N &0
\end{pmatrix}
$. 

Next, if we consider a fraction of $n$ nodes coupled with other nodes through the repulsive feedback couplings and a fraction of $1-n$ nodes coupled without the repulsive links (referred to as the  regular nodes), for a globally coupled network, the coupling matrix for the regular nodes will be given by $F_2$ and the coupling matrix corresponding to the repulsive nodes will be given by $F_1$. Substituting them in the Eq.~\ref{master} leads to the following eigenvalues,
\begin{equation}\label{eqn:eig2}
    \lambda_{1,2}=1-\varepsilon n \pm \sqrt{\varepsilon^2 n^2 -\omega^2}, \:\: \lambda_{3,4}=1-\varepsilon \pm i\omega
\end{equation}
The real part of these eigenvalues (Eq.~\ref{eqn:eig2}) must be negative for the amplitude death state to occur, which provides us the conditions $n>1/\varepsilon$ and  $\varepsilon < {1+\omega^2}$.  Fig.~\ref{fig:para_diagram}(a)-(b) confirms a perfect match between the numerical results and theoretical predictions for both the cases. 

\begin{figure}[t!]
    \centering
    	\includegraphics[height=4cm,width=\linewidth]{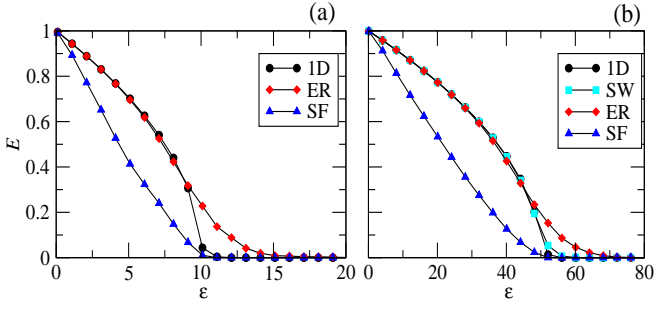}
    	\caption{$E$ vs $\varepsilon$, for non-identical oscillators with $\omega_0=5.0$  for various different network architectures (  black circle  scale-free, red triangle small-world, blue square  regular, and green triangle ER random networks). (a) $N=100$, $\langle k \rangle=10$ (b) $N=1000$, $\langle k \rangle=20$. }
    	\label{fig:model_networks}
\end{figure}

\paragraph{\bf{Dissimilar repulsive coupling in $x$ variable:}}
For the dissimilar repulsive coupling  applied only to the $x$ variable, the dynamics of the corresponding coupled equation will be governed by,
$\dot x_k= P^x_k
	-\frac{\varepsilon}{N}  \sum_{j=1}^{N}(y_j+x_k),\: \: \dot y_k=P^y_k$
The parameter space diagram obtained through the numerical calculations consists of AD and OS regions only. We find the necessary condition for these states to occur using the expression derived in Eq.~\ref{master}. The matrices required for the calculation are,
$
M=
\begin{pmatrix}
    \lambda-1+ \varepsilon &  \omega\\
    -\omega               & \lambda-1
\end{pmatrix}
\text{and }    
F=
\begin{pmatrix}
    0 & \varepsilon/N\\
    0 & 0
\end{pmatrix}
$
substituting them in Eq.~\ref{master} and solving this equation we get the following eigenvalues,
\begin{equation}
        \lambda_{1,2}= \frac{2- \varepsilon \pm \sqrt{\varepsilon^2-4\omega^2-4\omega\varepsilon}}{2},\:\:\lambda_{3,4}= \frac{2-\varepsilon \pm \sqrt{\varepsilon^2-4\omega^2}}{2}
        \label{eig_x}
\end{equation}

The necessary conditions can be derived from $Re[\lambda_{1,2}<0]$ (Eq.~\ref{eig_x}) as $\varepsilon>2$ and $\varepsilon<(1+\omega^2)/(1-\omega)$ while $\varepsilon^2-4\omega^2-4\omega\varepsilon>0$. From $Re[\lambda_{3,4}<0]$ (Eq.~\ref{eig_x}), one gets the necessary conditions as $\varepsilon>2$ and $\varepsilon>1+\omega^2$. However, the condition $2<\varepsilon<1+\omega^2$ prevails. This theoretical result matches with the numerical predictions illustrated in Fig.~\ref{fig:para_diagram}(c).
\\

\paragraph{\bf{ Dissimilar repulsive coupling through the $y$ variable:}}
Upon  applying the dissimilar repulsive coupling in only $y$ variable, the dynamical equation could be written as,
    \begin{equation}\label{eqn:SL_diss_y} 
    \dot x_k= P^x_k,\:\:\:\dot y_k=P^y_k-\frac{\varepsilon}{N}  \sum_{j=1}^{N}(x_j+y_k), \nonumber
    \end{equation}
 Again using the same procedure as in the last section, we analytically confirm a match with the numerical results.  
Here,
$
M=
\begin{pmatrix}
    \lambda-1   & \omega\\
    -\omega    & \lambda-1+\varepsilon
\end{pmatrix}
\text{and }    
F=
\begin{pmatrix}
    0 & 0\\
    \varepsilon/N & 0
\end{pmatrix}
$

Solving the generalized characteristic equation for these M and F values yields the following eigenvalues,
\begin{equation}
    \lambda_{1,2}= \frac{2- \varepsilon \pm \sqrt{\varepsilon^2-4\omega^2+4\omega\varepsilon}}{2}, \:\: \lambda_{3,4}= \frac{2-\varepsilon \pm \sqrt{\varepsilon^2-4\omega^2}}{2}
    \label{eig_y}
\end{equation}
For the origin to be stable, we need $Re[\lambda] < 0$. Therefore, from $Re[\lambda_{1,2}]<0$ (Eq.~\ref{eig_y}) we derive the conditions
$\varepsilon>2$ and $\varepsilon<(1+\omega^2)/(1+\omega)$, and similarly $Re[\lambda_{1,2}]<0$ (Eq.~\ref{eig_y}) yields $\varepsilon>2$ and $\varepsilon<(1+\omega^2)$. The condition $\varepsilon<(1+\omega^2)/(1+\omega)$ is dominant and provides us the governing equation characterizing the transition between the AD and OD states, which also matches with the numerical results (Fig.~\ref{fig:para_diagram}(d)).
\\

\paragraph{\bf{Non-identical oscillators:}}
Next, we consider the case of non-identical oscillators, i.e., $\omega_i \ne \omega_j$. For Lorentzian intrinsic frequency distribution given by, $g(\omega)=\frac{\Delta}{\pi[(\omega-\omega_o)^2+\Delta^2]}$, one obtain the following characteristic equation:
\begin{equation}
\begin{split}
     \frac{1}{\varepsilon^2}=
     &\Bigg[\int_{-\infty}^{+\infty}\frac{\lambda-1+\varepsilon}{(\lambda-1+\varepsilon)^2+(\omega)^2} g(\omega) d\omega)\Bigg]^2 \\
     &+\Bigg[\int_{-\infty}^{+\infty}\frac{\omega}{(\lambda-1+\varepsilon)^2+(\omega)^2} g(\omega) d\omega)\Bigg]^2
     \nonumber
\end{split}
\end{equation}
which can further be written as,\\
      $\frac{1}{\varepsilon^2}=\int_{-\infty}^{+\infty}\frac{1}{\lambda-1+\varepsilon+i\omega} g(\omega) d\omega \times\int_{-\infty}^{+\infty}\frac{1}{(\lambda-1+\varepsilon)-i(\omega)} g(\omega) d\omega$
The eigenvalues of this equation then can be given by,
\begin{equation}
    \lambda=1-\varepsilon+\Delta+\sqrt{\varepsilon^2-\omega^2,}
    \nonumber
\end{equation}
which yields the following condition for stability of the origin;
\begin{equation}\label{eqn:condition_dist}
   \text{if  } \Delta > 1,  \varepsilon >1;
   \text{ and 
   if }
   \Delta < 1 \text{, } 1 < \varepsilon <\frac{(1-\Delta)^2+\omega_o^2}{2(1-\Delta)}
\end{equation}
As seen from Fig.~\ref{fig: bif and non}(b), the numerical results are in agreement with the analytical results ( Eq.~\ref{eqn:condition_dist}). The time series of two nodes for a system($N=1000$) with Lorentzian distribution ($\omega_o=2.0$ and $\Delta=0.33$) is shown in Fig.~\ref{fig:timeseries}.
\\

\paragraph{\bf{Bifurcation diagram:}}
Fig.~\ref{fig: bif and non}(a) is drawn using XPPAUTO software \cite{Xppaut} depicting two types of bifurcation. The first one is reverse Hopf bifurcation (HB) where a stable origin transforms into an unstable origin along with two stable limit cycles as the coupling strength decreases. The second one is pitchfork bifurcation (PB) where a stable origin becomes unstable and two more symmetric fixed points come into existence as coupling strength increases.

\section{Numerical results for various model networks}
Next, we analyze the dynamical behaviors of coupled SL oscillators for this setup on various network architectures. We investigate how a  particular network structure affects the onset of the oscillator death by considering four different network architectures apart from the globally coupled network, namely, the regular  1-d lattice, ER random, small-world, and scale-free networks.

Among these, the scale-free and $1$-d lattice have the same lowest critical coupling strength at which AD occurs. While the small-world network has a slightly more critical coupling value as compared to the $1$-d lattice, ER random network achieves AD at higher critical coupling strength (Fig.~\ref{fig:model_networks}). The above observation implicates that the critical coupling strength increases when the regular $1$-d lattice is distorted and changed to the  ER random networks. Moreover,  with an increase in the average degree of these, we observe a similar rise in the critical coupling network (as shown in the supplementary material \cite{SM}).    

\section{Seizure data networks}
  \begin{figure}[t!]
    \centering
    	\includegraphics[height=6cm,width=0.5\textwidth]{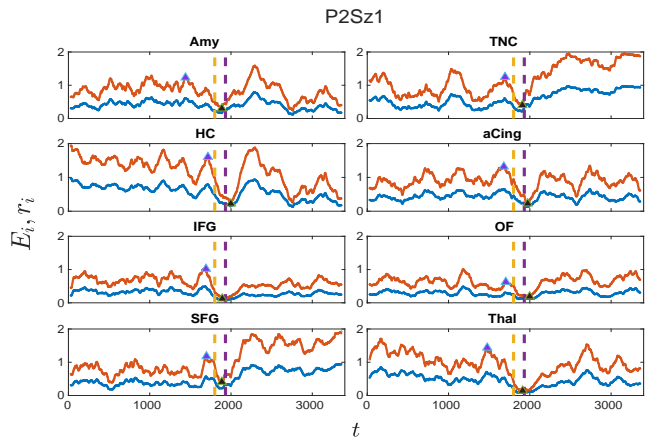}
    	\caption{{$\Delta$ band for  P2Sz1:} The red and blue lines represent $E_i$ vs $t$ and $r_i$ vs $t$,  respectively. The yellow and violet dashed lines correspond to  start and end of the ictal region. Each subfigure represents the dynamics of a node (full names are provided in the supplementary material. \cite{SM})  }
    \label{fig:P2sz1}    
\end{figure}  
 
 \begin{figure}[t!]
    \centering
    	\includegraphics[height=6cm,width=0.5\textwidth]{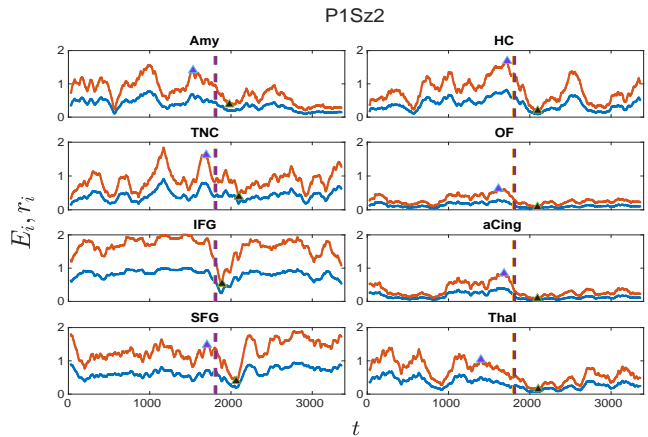}
    	\caption{$\Delta$ band for P1Sz2: Red and  blue lines represent $E_i$ vs $t$ and $r_i$ vs $t$, respectively. Yellow and violet dashed lines correspond to the beginning and  ending of the ictal region. Each subfigure represents dynamics of a node (full names provided in the supplementary material. \cite{SM}) }
    \label{fig:P1sz2}    
\end{figure}
Postictal generalized EEG suppression (PGES) refers to the diffuse background attenuation ($<10\mu V$) in the postictal state. The phenomenon is often observed following bilateral tonic-clonic seizures (TCS), and has been associated with a sudden unexpected death in epilepsy \cite{marjan_postic,Lhatoo_postic}. The mechanism and origin of occurrence of PGES are under intense investigations \cite{seyal,Kanth2022,isamu}. 
Here we show similarities of the phenomenon observed by Eq.~3 with PGES in convulsive seizure. We do not claim that the Eq.~3 presents an accurate model to the brain activities, nevertheless, the quenching of oscillations manifested by the model bears a resemblance to PGES. Moreover, to bring the model a step closer to the Brain activities, we present result for the coupling architecture corresponding to the correlations matrices for the EEG multivariate time series data for seizure. 
This correlation matrix dataset consists of $8 \times 8 \times f \times t$ tensor where $f$ and $t$ are, respectively, the number of the frequency levels and time steps for which data is recorded. The whole frequency range is divided into five bands or levels. The bands are as follows- $\Delta:2-4$, $\theta:4-7$, $\alpha:8-12$, $\beta:12-30$, $\gamma:30-40$.
Hence, each $8 \times 8$ matrix represents an adjacency matrix at a particular time at a particular frequency level. Each of these adjacency matrices is constructed by calculating the correlation between the time series of the 8 channels at the corresponding frequency level for a particular time window.
The detailed method for calculating gPDC matrices is given in supplementary material \cite{SM}.
\\

\paragraph{\bf Overall oscillation suppression:}
While, different regions for different patients reflect different behaviors, one typical pattern common in most of the nodes for all the patients is that the amplitude starts to decrease in the ictal region as compared to the preictal region. 
Ergo,  there exists a considerable suppression of the oscillation in the majority of the nodes around the ictal and in the initial stages of the postictal region for all the patients' data we have investigated. The amplitude remains low for the initial points of the postictal region and then slowly recovers with time.

\paragraph{\bf {Mechanism of  the amplitude death:}}
The uncoupled equation represents a limit cycle oscillator, to which the coupling acts as a decaying term resulting in the amplitude death at certain values of the frequency and the coupling strength. The impact of this coupling term as a whole also depends on the associated coupling matrices (underlying network architectures), and therefore there exists a change in the critical coupling strength for which death occurs  with the network structure. Additionally, the average degree of the network also plays a decisive role in deciding the amount of suppression of oscillations on the network. 

\section{Conclusion}
This article proposes a coupling setup that yields oscillation death in coupled Stuart-landau oscillators, and develops a theoretical framework to derive the necessary and sufficient conditions for attaining the oscillator death state for this setup with a fraction of nodes having repulsive feedback couplings. The analytical predictions are confirmed with the numerical experiments. Additionally, numerical results for the amplitude death on a few other model networks has been presented. Furthermore, we numerically analyzed the coupled dynamics model for the weighted correlations matrix constructed from the seizure data, and found that the phenomenon of amplitude suppression in the model resemble with the PGES. 
One of the immediate future extensions of this work is to derive analytical conditions for other states than AD, and to develop a generalized theoretical framework that can incorporate various forms of the couplings \cite{net_rev}, and adaptation rules \cite{SJ_NJP,SJ_PRE}. Further future directions are to have a more realistic model for the brain and to replicate these results in a larger cohort, particularly by including a postictal state that lacked PGES to understand the origin of PGES. 
Furthermore, the neural underpinning of such OD to AD  transition or oscillation suppression could also be harnessed towards developing neuro-modulation therapy principled to perturbation of the coupling process to prevent or rescue OD/AD.

\begin{acknowledgments}
SJ acknowledges support from the SERB POWER grant SPF/2021/000136.
\end{acknowledgments}

\end{document}